\documentclass[12pt]{article}
\usepackage{graphics}

\title{Energy and momentum losses in the  process of neutrino  scattering
on plasma electrons with the presence of a magnetic field.}

\author{Nickolay V. Mikheev \thanks{E-mail address:mikheev@univ.uniyar.ac.ru}
and
Elena N. Narynskaya \thanks{E-mail address:elenan@univ.uniyar.ac.ru}\\
{\small\it Division of Theoretical Physics, Department of Physics,}\\
{\small\it Yaroslavl State University, Sovietskaya 14,}\\
{\small\it 150000 Yaroslavl, Russian Federation.}}

\date{}

\begin{document}

\def\beq{\begin{equation}}
\def\eeq{\end{equation}}
\def\bd{\begin{displaymath}}
\def\ed{\end{displaymath}}
\def\nue{\nu \, e^-  \to \nu \, e^- }

\maketitle

\begin{abstract}

The neutrino-electron scattering in a dense dege\-ne\-rate
magnetized plasma under the conditions $\mu^2 > 2eB \gg \mu E$
 is inves\-ti\-ga\-ted.
 The volume density of the neutrino energy and
momentum losses due to this process are calculated. The results we
have obtained demonstrate that  plasma in the presence  of an
external magnetic field is  more transparent for neutrino than
non-magnetized plasma. It is shown that neutrino scattering under
conditions considered does not lead to the neutrino force acting
on plasma.
\\
\\
{{\bf Key words:} neutrino-electron processes, plasma, magnetic
field, supernova envelope}
\\
\\
PACS numbers: 13.15.+g, 95.30.Cq, 97.60.Bw

\end{abstract}

\newpage
\section{Introduction.}

\indent

It is known that the neutrino physics plays an unique  role in
astro\-phy\-sics and cosmology.
 In particular, these
light weakly interacting  particles are special for astrophysical
phenomena like a supernova explosion when  a large number of
neutrinos is produced in a collapsing stellar core (Raffelt 1996). The compact
core  with the typical radius $R \sim 10 \, $km, the supranuclear
density $\rho \sim 10^{14} \, g/cm^3$ and the high temperature $T
\sim 30 $ MeV, is opaque for neutrinos. While a rather rarified
remnant envelope with the typical density   $\rho \sim 10^{10} -
10^{12} \, g/cm^3$ and temperature of the order of few MeV,
becomes partially transparent for the neutrino flux.

Notice that in investigations of neutrino processes in  medium not
only  dense substance, but also a magnetic field should be taken
into account. We stress that a magnetic field can play the role of
additional component of an active medium, and influence
substantially on particle properties. This influence becomes
especially important in the case when the magnetic field strength
reaches the critical, Schwinger value $B_e = m^2_e/e = 4.41 \times
10^{14}$ G.~\footnote{ We use natural units in which $c=\hbar=1$,
$e > 0$ is the elementary charge. } According to modern
astrophysical models, very strong magnetic fields up to $10^{17}$
G could be generated, for example, in a rapidly rotating supernova
remnant (Duncan and Thompson 1992;
Bisnovatyi-Kogan 1993; Mathews et.al. 1997).

Previously, in the studies of neutrino interactions with a dense
stellar medium the main attention was given to the neutrino --
nucleon processes. This is due to the  fact that the
Urca-processes and the  neutrino - nucleon scattering defined the
major contribution into the energy balance of the col\-lap\-sing
 core,  and were considered as a  main source of neutrino opacity.
The neutrino - electron processes were less investigated. However,
as it was pointed in studies Mezzacappa and Bruenn (1993),
 taking  account  of the
neutrino - electron scattering in a detail ana\-ly\-sis of the
supernova dynamics  is physically justified indeed. In particular,
the neutrino - electron processes can contribute significantly
into the asymmetry and  provide a competition with the
neutrino-nucleon processes. For example,
 in the paper Kuznetsov and Mikheev (2000) the total set of
neutrino-electron processes ($\nu e^\pm \leftrightarrow \nu
e^\pm$, $ \nu e^- e^+ \leftrightarrow \nu$, $ e^\pm
\leftrightarrow \nu \bar \nu e^\pm$) was investigated in a strong
magnetic field limit, when electrons and positrons occupied the
lowest Landau level. It was shown that the neutrino force action
on plasma along the magnetic field  turns out  to be of the same
order  and, what is essential, of the same sign
 as the one caused by the $\beta $-processes (Gvozdev and Ognev 1999).

By this means,  the  investigations
of the neutrino-electron processes  under extreme conditions of a
high density and/or  tem\-pe\-ra\-ture of matter and also of a
strong magnetic field are the subject of a great interest.

In this study we investigate the neutrino -- electron  processes
in a dense magnetized plasma. In contrast to (Kuznetsov and
Mikheev  2000) we consider the physical situation when the
magnetic field is not so strong, whereas the density of plasma is
large. Thus the chemical potential of electrons, $\mu$, is the
dominating factor:
\beq \mu^2 > 2eB \gg T^2, E^2 \gg m_e^2, \label{eq:cond1} \eeq
where $T$ is the plasma temperature, $E$ is the typical neutrino
energy. Under  the conditions  (\ref{eq:cond1}) plasma electrons
occupy the excited Landau levels. At the same time it is assumed
that the magnetic field strength being relatively weak,
(\ref{eq:cond1}), is simultaneously strong enough, so that the
following condition is satisfied:
\beq
 eB \gg \mu E.
  \label{eq:cond2}
\eeq
In the present astrophysical view, the conditions
(\ref{eq:cond1}), (\ref{eq:cond2}) could be realized, as an
example, in a supernova envelope, where the electron chemical
potential is assumed to be $\mu \sim 15$ MeV, plasma temperature
$T \sim 3$ MeV. The magnetic field could be as high as $10^{15} -
10^{16}$ G. Under the conditions considered the approximation of
ultrarelativistic plasma is a good  one, so we will neglect the
electron mass wherever this causes no complications.

As it was shown in paper Mikheev and Narynskaya (2000), under the
conditions (\ref{eq:cond1}), (\ref{eq:cond2})
 the total set of neutrino -
electron processes reduces to the process of neutrino scattering
on plasma  electrons.  Moreover, both initial and final electrons
occupy the same Landau level.

The neutrino -- electron scattering in dense magnetized plasma was
in\-ves\-ti\-ga\-ted by  Bezchastnov and Haensel (1996).
Numerical calculations of the differential cross -- section of
this process in the limit of a weak magnetic field ($eB < \mu E$)
were performed. The purpose of our work is to calculate
analytically
 not only  the probability of this process, but also the
volume density of the neutrino energy and momentum losses under
the conditions (\ref{eq:cond1}),(\ref{eq:cond2}).

\section{Probability of the neutrino-electron scat\-te\-ring.}

\indent

We start from the effective local Lagrangian of the neutrino --
electron interaction in the framework of  the Standard Model:
\beq
 L_{eff}= \frac{G_F}{\sqrt2}
  [\overline e \gamma_{\alpha}(c_v - c_a\gamma_5)e] \, j^{\alpha},
\label{eq:lag}
\eeq
where $ j^{\alpha} = [\overline \nu\gamma_{\alpha}(1 -
\gamma_5)\nu]$ is the current of  massless left neutrinos,
$c_v=\pm1/2 + 2sin^2\theta_W$, $c_a=\pm 1/2$. Here upper signs
correspond to the electron neutrino $(\nu = \nu_e)$ when both Z
and W boson exchange takes part in a process. The lower signs
correspond to $\mu$ and $\tau$ neutrino $(\nu =
\nu_\mu,\nu_\tau)$, when the Z boson exchange is only presented in
the Lagrangian (\ref{eq:lag}).

In order to impart a physical meaning to the probability of the
neutrino - electron scattering per unit time, it is necessary to
integrate not only over the final but also over the initial
electron states as well:
\begin{equation}
W_{(\nu e^- \to \nu e^-)}= \sum_{n=0}^{n_{max}}
\frac{1}{{\cal T}} \int  \sum_{s,s'}
\mid S\mid^2 \,
 dn_{e^-} \, dn'_{e^-}  \,
\frac{d^3k'}{(2\pi)^3} \, V  \, (1-  f(E')).
 \label{eq:defprob}
\end{equation}
Here $n_{max}$ corresponds to the maximal possible Landau level
number, which  is defined as the integer part of the ratio $ \mu^2
/2eB \ge 1 $, ${\cal T}$ is the total interaction time, $\mid S
\mid ^2$ is  the S-matrix element squared of the process
considered, $V$ is the normalization volume,
 $f(E')$ is a distribution function of final neutrinos,
$  f(E') = [e^{(E' - \mu_\nu)/T_\nu} + 1]^{-1}$, $E'$ is the
final neutrino  energy,
 $\mu_\nu$ and $T_\nu$ are the effective chemical potential and
the spectral  temperature of the neutrino  gas correspondingly. In
a general case the neutrino spectral temperature $T_\nu$ can
differ from the plasma temperature $T$ (we do not assume the
equilibrium between
 neutrino  gas and plasma).
The phase-space elements of the initial and final plasma electrons in
the presence of a magnetic field are defined by the following way:
\footnote{ we use the  gauge $A^\mu =(0,0,Bx,0)$, the magnetic
field is directed along the $z$ axis.}
\bd
 dn_{e^-} = \frac{dp_y \,dp_z}{(2\pi)^2 } \,L_y \,L_z \,
f(\varepsilon_n),
 \,\,\,\,
 dn'_{e^-} = \frac{dp'_y \,dp'_z}{(2\pi)^2 } \,L_y \,L_z \,
(1 - f(\varepsilon'_n)),
\ed
 where $p_z$ is the electron momentum along the magnetic
field, $p_y$ is the generalized  momentum  which defines the
position of the center of a Gaussian packet along the $x$ axis,
$x_0=-p_y/eB $,  while $\varepsilon_n \simeq \sqrt{p_z^2 + 2eBn}$
is the energy of  an ultrarelativistic plasma electron occupying
the $n$-th Landau level,
$f(\varepsilon_n)$ is a distribution function of electrons,
$f(\varepsilon_n) = [e^{(\varepsilon_n - \mu)/T} + 1]^{-1}$.

The details of integration over the phase space of particles had
been published in our previous paper (Mikheev and Narynskaya 2000). The
result of calculation of the probability (\ref{eq:defprob}) can be
presented in the relatively simple form:
\begin{eqnarray}
  W_{(\nue)}  & = & \frac{G_F^2\,(c_v^2+c_a^2) \,eB \,T^2\, E}{4\,\pi^3}
 \,   \sum_{n=0}^{n_{max}} \, \frac{1}{z^2}
 \nonumber
\\
&\times & \bigg \lbrace
  ((1+z^2)(1+u^2)-4uz)  \int \limits_{-a}^{b}  \Phi (\xi) \,d\xi
\label{eq:prob1}
\\
 & + & \frac{1}{zr \tau}(z^2 -1)(z-u)  \int \limits_{-a}^{b}  \xi \,
   \Phi (\xi)\, d\xi \,\,\,  \bigg \rbrace
\,\,\,\,  +  \,\,\,\, ( u \to - \,u ), \nonumber
\end{eqnarray}
where $z=\sqrt{1-2eBn/\mu^2}$, $\Phi
(\xi)=\xi[(e^\xi-1)(e^{\eta_\nu-r-\xi/\tau}+1)]^{-1}$, $a=r\tau
z(1+u)/(1+z)$ and  $b=r\tau z(1-u)/(1-z)$, $ r=E/T_\nu$, $\tau
=T_\nu/T$, $\eta_\nu=\mu_\nu/T_\nu$, $u=cos\theta$, $\theta$ is
the angle between the initial neutrino
 momentum $\vec k$ and the magnetic field direction. The variable
 $\xi$ defines the spectrum of the probability (\ref{eq:prob1})
 on the final neutrino energy, $\xi = (E'- E)/T$.

In the limit of a very dense plasma  $(\mu^2 \gg eB)$, when  a
great number of Landau levels are occupied by plasma electrons,
one can change the summation over $n$  by integration over $z$:
\bd \sum_{n=0}^{[\mu^2/2eB]} \, F(z) \simeq \frac{\mu^2}{eB}
\int\limits_0^1 F(z) \, z \,dz. \ed
In this case the contribution from the lowest Landau levels turns
out to be negligibly small, so the main contribution into the
probability arises from the highest Landau levels. In this limit
the probability (\ref{eq:prob1}) can be rewritten in the following
form:
\begin{eqnarray}
  W_{(\nue)} & = & \frac{G_F^2(c_v^2+c_a^2)\, \mu^2 \,T^2 E}{4\,\pi^3}
   \int\limits_0^1  \frac{dz}{z} \,
   \nonumber\\
  &   \times &  \bigg \lbrace
  ((1+z^2)(1+u^2)-4uz)\int \limits_{-a}^{b} \Phi (\xi) d\xi
\label{eq:prob2} \\
  & + &  \frac{1}{zr \tau}(z^2 - 1)(z-u)  \int \limits_{-a}^{b}  \xi
   \Phi (\xi) d\xi  \,\,\, \bigg \rbrace  + \,\,\,  ( u \to - \,u ).
   \nonumber
\end{eqnarray}
As one can see, the probability (\ref{eq:prob2}) does not depend
on the value of the magnetic field strength, but is not isotropic.
The dependence on the angle $\theta$ manifests this anisotropy of
the neutrino-electron process in the presence of a magnetic field.
In the limit of  a rare neutrino gas when $f(E') \ll 1$,  the
result has a more simple form:
\beq W_{(\nue)} \simeq \frac{G^2_F(c_v^2+c_a^2)\mu^2 E^3}{12\pi^3}
\, I(u) , \label{eq:prob3} \eeq
\bd
 I(u) =  \int\limits_0^1 \frac{z dz}{(1 + z)^2}  \,
(u^4 \, (3 z^2 + 2z +1) - 12 u^2 z + z^2 +2z +3).
\ed
For
comparison we present here the probability of the neutrino -
electron scattering  without field in the same limitof the rare
neutrino gas :
\beq
 W_{vac}=\frac{G^2_F(c_v^2+c_a^2)\mu^2E^3}{15\pi^3}.
 \label{eq:probvac}
\eeq
\begin{figure}[t]
\centerline{\includegraphics{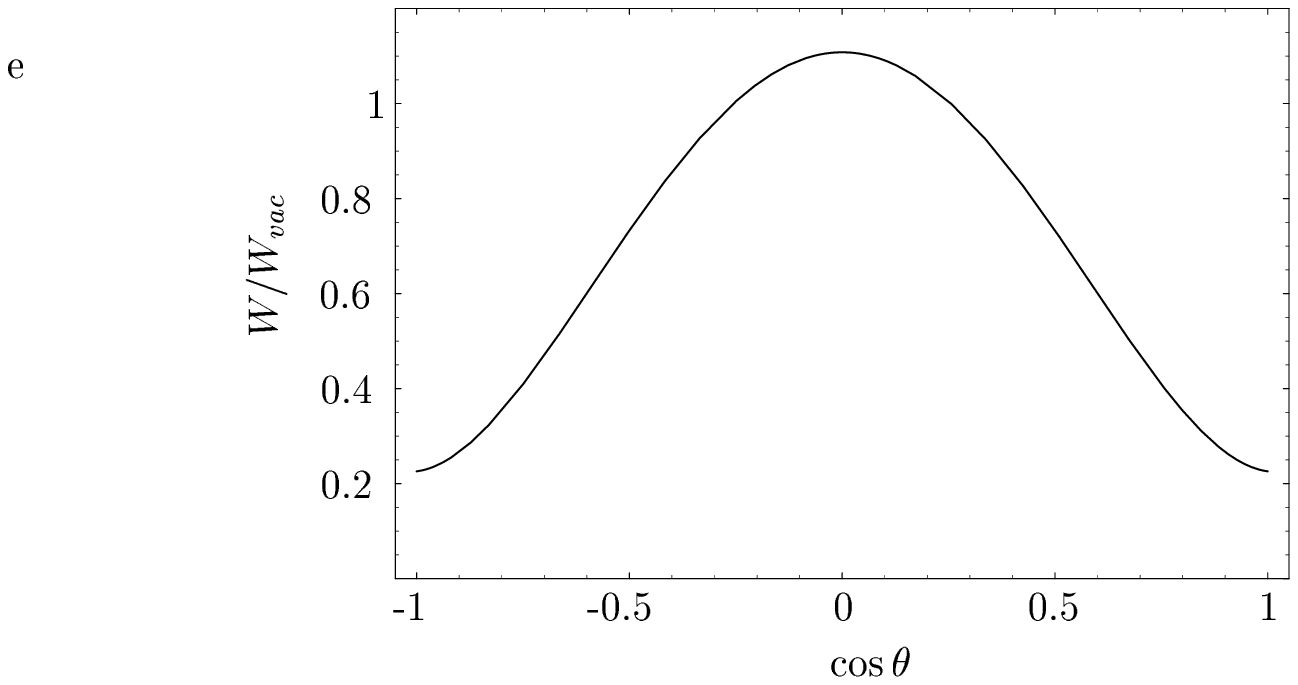}} \caption{The
relative probability of the neutrino-electron scattering in a magnetized plasma
as a function of the  angle between the initial neutrino
 momentum  and the magnetic field direction. $W_{vac}$ is the
 probability in a non-magnetized plasma.}
\label{fig:graf1}
\end{figure}
The numerical estimation of the ratio of the probabilities
(\ref{eq:prob3}) and (\ref{eq:probvac}) is presented in
Fig.\ref{fig:graf1}. It is seen that  the probability in a
magnetized plasma excesses the vacuum probability
 in the  vicinity of a point $\theta=\pi/2$ only.

\section{Integral  neutrino action on a magnetized plas\-ma.}

\indent

In this section we will calculate the volume density of neutrino
energy and  momen\-tum losses  per unit time in a medium, which
could be defined by the following way:
\beq
  (\dot \varepsilon, \vec{\cal F}) =
       \frac{1}{(2\pi)^3}\int \frac{(q_{0}, \, \vec
       q)\,d^3k}{e^{(E-\mu_\nu)/T_\nu}+1}\,dW,
       \label{eq:defloss}
\eeq
where $q_\alpha$ is the difference between the momenta of the
initial and final neutrinos, $q_\alpha = k_\alpha - k'_\alpha$. The
zeroth component, $\dot \varepsilon$, determines the neutrino
energy loss in unit volume per unit time. In general, a neutrino
propagating through plasma can both lose and capture  energy.
So, we will mean  the ''loss'' of energy in the algebraic
sense.

The vector $\vec{\cal F}$ in Eq.(\ref{eq:defloss}) is associated with
the volume density  of the neutrino  momentum loss in unit time,
and therefore it defines the neutrino force acting on plasma.
Because of the isotropy of plasma  without a magnetic field, in
the presence of a magnetic field one would expect to obtain the
neutrino force action along the magnetic field only. However, as
it was shown above, the probability of the neutrino-electron
scattering (\ref{eq:prob2}) is the symmetric functions with
respect to the substitution $u \to - u$ (or $\theta \to \pi -
\theta$). This means that the neutrino scattering on excited
electrons does not give a contribution into the neutrino force
acting on plasma along the magnetic field. Thus, under the
conditions (\ref{eq:cond1}), (\ref{eq:cond2}) there is no
neutrino force action on plasma at all. Therefore, this force is
caused by a contribution of neutrino interactions with  ground
Landau level electrons only, and the result obtained  by
Kuznetsov and Mikheev (2000) has a more general applicability in
fact. It may be used even in the  limit of dense
 plasma when
chemical potential is considerably greater than the magnetic field
strength ($\mu^2 \gg e B$).

For the  neutrino  energy loss in unit volume per unit time in the
limit of a very dense plasma we obtain the following result:
\beq
\dot \varepsilon_{B} =
 \frac{G^2_F  (c_v^2 + c_a^2)}{\pi^3} \,\mu^2 \, T^4 \, n_\nu
       \, J_{B}(\tau), \label{eq:lossB}
 \eeq
\begin{eqnarray}
 J_{B}(\tau) &  = & \frac{\tau^4}{2} \, \int\limits_0^1 \frac{dz}{z^2} \int\limits^\infty_0 dy \, y^2
 \,[ y (1  -  z^2) + \, 4\, z\, (1 + z^2)\,]\nonumber
 \\
 & \qquad &   \hspace{30mm}  \times \,\, \frac{1 - e^{y(1 - \tau)}}{1 -
e^{-y\tau}} \,\, e^{-y (1+z)/ 2z},  \label{eq:JB}
\end{eqnarray}
where $n_\nu$ is the concentration of initial neutrinos, the
parameter $\tau$ has a meaning of a relative  neutrino spectral
temperature, $\tau = T_\nu/T$. It is interesting to compare this
result with the one in a non-magnetized plasma which can be
presented in a similar form:
\beq \dot \varepsilon_{B=0}  =   \frac{G^2_F  (c_v^2 +
c_a^2)}{\pi^3} \,\mu^2 \, T^4 \, n_\nu \, J_{B=0}(\tau),
\label{eq:loss0}\eeq
\beq
 J_{B=0}(\tau)   =  4 \tau^4\, \int\limits_0^\infty \, d\xi
\,\xi^2 \,
 \frac{e^{\xi(\tau - 1)} - 1}{e^{\xi\tau} - 1}.
 \label{eq:J0}
\eeq
The functions  $J_{B}(\tau)$  and $J_{B=0}(\tau)$ define the
dependence of the neutrino energy losses on the relative neutrino
spectral temperature in a magnetized plasma and in a plasma
without field correspondingly. In the limit of a sufficiently
large neutrino spectral temperature ($e^{\tau} \gg 1$) they reduce
to the power functions:
$$ J_{B}(\tau) \simeq 4.35 \,\, \tau^4, \,\,\,J_{B=0}(\tau) \simeq 8 \,\,\tau^4.$$
\begin{figure}[t]
\centerline{\includegraphics{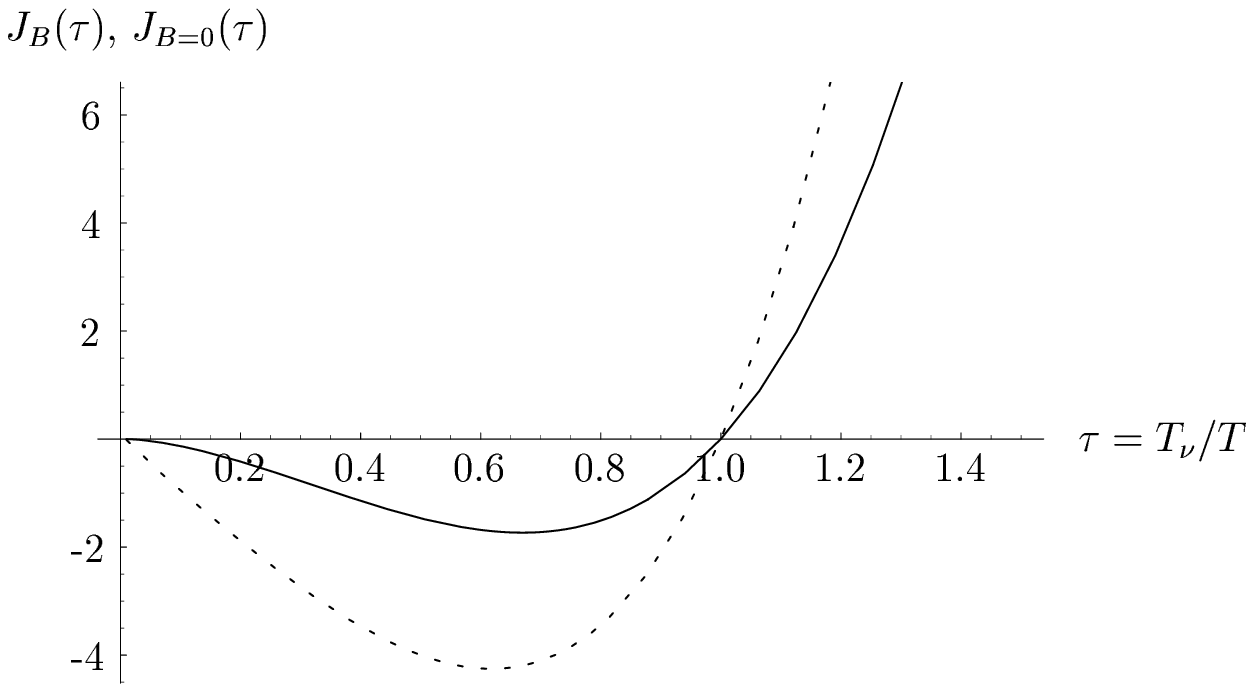}} \caption{The  functions
$J_{B}(\tau)$ (solid line) and $J_{B=0}(\tau)$ (dashed line)
versus  the relative spectral neutrino temperature.}
\label{fig:graf2}
\end{figure}

The graphs of the functions (\ref{eq:JB}) and (\ref{eq:J0})
 are presented in
Fig.\ref{fig:graf2}. As one would expect,
 at  neutrino spectral temperature smaller than the plasma one ($\tau < 1$)
the values of the functions $J_{B}(\tau)$  and $J_{B=0}(\tau)$ are
negative. It implies that  neutrino propagating via  medium
captures energy from the plasma. At the $T_\nu $ more then $T$
($\tau > 1$),  neutrino
gives up energy to the plasma.
 At the point $\tau=1$
there is a thermal equilibrium  when there is no energy exchange
between neutrino and electron-positron plasma. It can be seen that
the neutrino energy loss in a magnetized plasma  is less than the
one in non-magnetized plasma.  By this means,
 under the  conditions (\ref{eq:cond1}), (\ref{eq:cond2})
 the magnetized plasma becomes more transparent for
neutrinos than  plasma without field.

\section{Conclusions.}

\indent

In this paper we have investigated the neutrino-electron
scattering in a dense magnetized plasma. We have considered the
physical situation when
the plasma component is the dominating one
of the two components of the active medium.
 At
the same time, a magnetic field was assumed to be not too small
($\mu^2
> 2eB \gg \mu E$ ).  The probability and the  volume density of the
neutrino energy-momentum losses have been calculated.

It is found, that the neutrino scattering on excited electrons
does not give a contribution into the neutrino force acting on
plasma. This force is caused by the neutrino-electron processes
when plasma electrons occupy the lowest Landau level only. Thus
the result  for this force, obtained in paper Kuznetsov and
Mikheev (2000), has a more wide  area of application. It may be
used under a condition $\mu^2 \gg eB$ as well.

It is shown that under the  conditions  (\ref{eq:cond1}),
(\ref{eq:cond2}) the combine effect of plasma and strong magnetic
field leads to a decrease of the neutrino energy loss in
comparison to the one in a pure plasma. Therefore, the complex
medium, plasma + strong magnetic field, is more transparent for
neutrinos than non-magnetized plasma.

One would believe  that the result obtained will be useful
 for the detail analysis of  astrophysical cataclysms like
 supernova explosions.

\vspace{10mm}

\section*{Acknowledgments}

This work was supported in part by the Russian Foundation for
Basic Research under the Grant No. 01-02-17334 and  by the
Ministry of Education of Russian Federation under the Grant No.
E00-11.0-5.


\addcontentsline{toc}{part}{\bf \refname}
\begin{thebibliography}{20}

\bibitem{Bezchastnov}
   Bezchastnov, V.G. and Haensel, P., 1996.
   Neutrino-electron scattering in a dense magnetized plasma.
    {\sl  Physical Review D}, 54 (6), 3706-3721.

\bibitem{Bisnovatyi}
   Bisnovatyi-Kogan, G.S., 1993.
   {\sl  Physical Problems of Theory of Stellar Evolution}.
   Nauka: Moscow.

\bibitem{Duncan}
  Duncan, R.C. and Thompson, C., 1992.
  Formation  of very strongly magnetized neutron stars: implications for
  gamma-ray bursts.
   {\sl Astrophysical Journal},  392, L9-L13.

\bibitem{Gvozdev}
  Gvozdev, A.A. and  Ognev, I.S., 1999.
 On  the possible enhancement of the magnetic field by neutrino
 reemission processes in the mantle of a supernova.
  {\sl JETP Letters},  69 (5), 365-370.


\bibitem{Mikheev}
    Kuznetsov, A.V. and  Mikheev, N.V., 2000.
    Neutrino interaction with strongly magnetized electron-positron plasma.
     {\it Journal of Experimental and Theoretical Physics},
      91 (4), 748-760.

\bibitem{Mathews}
   Mathews, G.J. et. al., 1997.
    Gamma-ray bursts from neutron star binaries.
  Preprint astroph/9710229.

\bibitem{Mezzacappa}
   Mezzacappa, A. and  Bruenn, S.W., 1993.
    Stellar core collapse: a boltzmann treatment of  neutrino-electron
   scattering.
    {\sl Astrophysical Journal}, 410, 740-760.


\bibitem{Narynskaya}
   Mikheev, N.V. and Narynskaya, E.N., 2000.
   Neutrino-electron processes in a dense magnetized plasma.
   {\sl Modern Physics Letters A}, 15 (25), 1551-1556.

\bibitem{Raffelt}
  Raffelt, G.G., 1996.
   {\sl  Stars as Laboratories for Fundamental Physics}.
  University of Chicago Press: Chicago.




\end{thebibliography}
\end{document}